\documentclass[a4paper,fleqn]{cas-sc}
 
\usepackage[numbers,sort&compress]{natbib}
 
\usepackage{amsmath,amssymb}
\usepackage{bm}
\usepackage{siunitx}
\usepackage{xcolor}
\usepackage{comment}
 
\def\tsc#1{\csdef{#1}{\textsc{\lowercase{#1}}\xspace}}
\tsc{PVA}
\tsc{XRD}
 
\begin{document}
\let\WriteBookmarks\relax
\def\floatpagepagefraction{1}
\def\textpagefraction{.001}
 
\shorttitle{Polymer-regulated freezing of water droplets}
 
\shortauthors{H. An et~al.}
 
\title[mode=title]{Polymer-Regulated Freezing of Water Droplets Revealed by Synchrotron X-ray Imaging and Raman Spectroscopy}
 
\author[1]{Hyeonjun An}
[orcid=0009-0000-5316-6264]
\credit{Writing -- original draft, Formal analysis, Data curation, Investigation, Conceptualization}
 
\author[2]{Bomi Kim}
\credit{Writing -- original draft, Formal analysis, Data curation, Investigation}

\author[1]{Jae Kwan Im}
\credit{Data curation, Investigation}

\author[3]{Min Woo Kim}
\credit{X-ray imaging experiment}
 
\author[3]{Seob-Gu Kim}
\credit{X-ray imaging experiment}

\author[3]{Jae-Hong Lim}
\credit{X-ray imaging experiment}
 
\author[2]{Kitae Kim}[orcid=0000-0002-3952-0287]
\cormark[1]
\ead{ktkim@kopri.re.kr}
\credit{Supervision, Project administration, Funding acquisition}

\author[1,4]{Joonwoo Jeong}[orcid=0000-0002-3082-783X]
\cormark[1]
\ead{jjeong@unist.ac.kr}
\credit{Writing -- original draft, Supervision, Project administration, Funding acquisition, Conceptualization}
 
\affiliation[1]{organization={Department of Physics, Ulsan National Institute of Science and Technology (UNIST)},
                city={Ulsan},
                country={Republic of Korea}}
 
\affiliation[2]{organization={Korea Polar Research Institute (KOPRI)},
                city={Incheon},
                country={Republic of Korea}}
 
\affiliation[3]{organization={Pohang Accelerator Laboratory, Pohang University of Science and Technology (POSTECH)},
                city={Pohang},
                country={Republic of Korea}}
 
\affiliation[4]{organization={UNIST Research Center for Soft and Living Matter, Ulsan National Institute of Science and Technology (UNIST)},
                city={Ulsan},
                country={Republic of Korea}}
 
\cortext[cor1]{Corresponding authors.}



\begin{abstract}
Adding a polymer to a sessile water droplet not only lowers its freezing point but also suppresses the tip singularity that forms during its freezing on cold substrates.
Here, we employ synchrotron X-ray and Raman imaging to elucidate the spatiotemporal mechanism underlying tip suppression in an aqueous polyvinyl alcohol (PVA) solution, a model polymer solution.
As the polymer concentration increases, we observe slower propagation of the freezing front, reduced bubble entrapment, and a progressively more rounded apex across the volumes and molecular weights examined.
X-ray tomography reveals that frozen PVA droplets retain low X-ray transmittance domains in their interiors and at the surface, and Raman spectral mapping confirms that these domains correspond to PVA-enriched regions, providing direct evidence of freeze-induced polymer segregation.
These findings indicate that PVA is redistributed heterogeneously during water solidification rather than shifting bulk properties homogeneously, providing a spatially resolved framework for interpreting the observed tip blunting and the suppression of discrete bubble entrapment.
Our work identifies freeze-induced polymer segregation as a pathway by which a dissolved polymer regulates both the external shape and the internal structure of a freezing droplet, and these findings shed light on potential applications in freezing-based processes such as freeze-casting and cryopreservation.
\end{abstract}

\maketitle
\section{Introduction}

The freezing of water drives phenomena ranging from climate dynamics in Arctic seawater to the engineered microstructures of freeze-cast materials. This phase transition is governed by a complex interplay of multi-physics processes, including supercooling, latent heat dissipation, interfacial evolution, volumetric expansion, gas rejection, and solute redistribution \cite{Chu2024SaltwaterNatComm, Zhu2025SaltwaterBurrLike, Wang2024SelfLiftingPRL, Shen2025EthanolDropletBulge}.
The freezing of a sessile water droplet provides an ideal model system for isolating and studying this complexity.
During solidification, the freezing front propagating from the droplet bottom adopts a self-similar spherical shape, while mass conservation and volumetric expansion focus the remaining liquid into a point, producing a characteristic conical apex \cite{Snoeijer2012PointyIceDrops, Marin2014TipSingularity}.
Simultaneously, dissolved gases rejected by the advancing ice front supersaturate the remaining liquid, leading to bubble nucleation, growth, and entrapment \cite{Chu2019BubbleFormation, Li2022DissolvedGas, Meijer2024BubbleGrowth, Shao2024BubbleDistribution, Shao2023BubbleGrowthDistribution}.
The resulting pore morphologies reflect the delicate interplay among freezing rate, gas diffusion, and capillarity \cite{Snels2023IceDropletXCT, Thievenaz2025BubbleShape, Im2026XrayMicroscopyDraft}.
Thus, the freezing of a neat water droplet has served as a reference for understanding how additives modify its phase transitions and morphology.

Polymer additives offer a means to regulate droplet freezing, and an aqueous polyvinyl alcohol (PVA) solution has been a representative model system. 
PVA, a water-soluble polymer, has been extensively studied for its effects on ice growth and ice recrystallization \cite{Budke2006PVAIRI, Naullage2018PinnedIceSurfaces, Naullage2020PVAIRI, Bachtiger2021PVAIRI}.
Prior studies have shown that PVA indeed alters freezing-front morphology, creating rough, needle-like, rippled, faceted, or tilted interfaces rather than the smoother fronts observed in neat water \cite{Weng2018PVAIceInterface, Zhang2021SingleIceCrystal, Zhang2021Lamellar, Piao2021PVAIRIneedleShape}.
At the droplet scale, increasing PVA concentration suppresses tip formation and prolongs total freezing time.
These phenomena have been explained by changes in the effective density ratio and concentration-dependent thermophysical properties~\cite{Kharal2024PolymerDroplets}.
Researchers have also noted that the slow diffusion of solute --- the polymer rejected from ice --- causes persistent concentration gradients to develop ahead of the advancing freezing interface \cite{Ulrich2026PolymerFreezingReview, Wegst2010BiomaterialsFreezeCasting, Yin2019FreezeCastChitosan, Yin2023PolymerFreezeCasting}. 
Even a universal scaling in freezing dynamics based on competition between viscous stresses, capillarity, and solute transport has been proposed~\cite{Ulrich2026UniversalScalingPreprint}.
Crucially, however, these approaches do not address where the rejected polymer accumulates within the droplet, how it is spatially organized during freezing, or how this dictates the evolving frozen internal structure.
Addressing these questions requires a probe that penetrates the polycrystalline ice obscuring the droplet interior, and X-ray imaging has proven well suited to characterizing a range of freezing and solidification processes \cite{Im2026XrayMicroscopyDraft,Kamm2023XraySucrose,Alpert2022XrayIceNucleation,Wu2025XraySaltwater,Jin2017XrayINP}.

Here, we bridge this gap by combining synchrotron X-ray and Raman imaging to simultaneously track the evolution of the freezing front, the internal structure, and the polymer redistribution in sessile PVA droplets. By directly resolving both morphology and chemical composition in situ, we reveal that macroscopic bulk properties are insufficient to explain the physics of polymer-modified freezing. Instead, we demonstrate that understanding polymer-regulated freezing fundamentally requires accounting for heterogeneous polymer redistribution during solidification.

\section{Materials and Methods}

\subsection*{Sample preparation}
We used deionized water of \SI{18.2}{\mega\ohm\centi\meter} for all experiments and the following polymers: polyvinyl alcohol ($M_\mathrm{w} =$ 13{,}000--23{,}000, 98\% hydrolyzed, and $M_\mathrm{w} =$ 85{,}000--124{,}000, 99+\% hydrolyzed; Sigma-Aldrich). 
The stock solutions of high polymer concentration, with no undissolved residue (e.g., 3~wt\%), were diluted to obtain solutions of the target concentrations.
A water-cooled and PID-controlled Peltier device controlled the temperature of an aluminum substrate from room temperature down to \SI{-15}{\degreeCelsius} under ambient conditions of \(21.7 \pm 0.6\)\,\si{\degreeCelsius} and a relative humidity of \(57.8 \pm 2.6\)\%.
After manually depositing a polymer-solution droplet of a controlled volume (\SIrange{1}{10}{\micro\liter}) onto the substrate at room temperature, we initiated X-ray imaging and acquired the first image within approximately \SI{15}{\second}. 
Then, the substrate was cooled to the target temperature at a rate of \SI{-2}{\degreeCelsius\per\second}; the onset of cooling was defined as the time at which the programmed substrate cooling began.

\subsection*{X-ray imaging setup and analysis}

We performed X-ray imaging experiments at the 6C Bio Medical Imaging beamline of Pohang Light Source-II (PLS-II).
A monochromatic  \SI{14}{\kilo\electronvolt} X-ray beam was used, and a 2 mm graphite filter was placed upstream to attenuate low-energy photons, thereby reducing unnecessary dose and mitigating the risk of beam-induced damage to the PVA solution. 
To assess potential X-ray damage to PVA during freezing and imaging, we compared X-ray tomography results for samples frozen with and without X-ray irradiation during freezing. We found no discernible difference, as shown in Fig. S3.

The effective pixel size and exposure time for a typical snapshot were \SI{0.65}{\micro\meter} and \SI{100}{\milli\second}, respectively; we recorded the freezing process at 10~frames per second.
Because of the safety procedure required for hutch evacuation and shutter opening, the time interval between sample placement and acquisition of the first image was approximately \SI{15}{\second}.

Flat-field correction was performed using a flat-field (open-beam) image acquired without the sample and a dark image acquired with the X-ray beam off to correct for detector dark noise~\cite{Seibert1998FlatField}.
The normalized image was calculated as $I_{\text{norm}} = \frac{I_{\text{sample}} - I_{\text{dark}}}{I_{\text{flat}} - I_{\text{dark}}}$.
Note that we adjusted the overall contrast of the figures in the manuscript only to enhance interfacial visibility, and the raw images are available from~\cite{Zenodo}.

For tomography of frozen specimens, the sample stage was rotated by \ang{180}, and 361 projection images were collected at \ang{0.5} intervals over \SI{36}{\second}.
We used Octopus reconstruction software (Ver. 8.7, XRE, Belgium) for three-dimensional reconstruction.

\subsection*{Raman imaging setup and analysis}
A \SI{10}{\micro\liter} PVA droplet was frozen on a temperature-controlled stage (THMS600, Linkam Scientific Instruments, UK) and then transferred to a cold room at \SI{-10}{\degreeCelsius} to minimize thawing and frost formation during sample preparation.
In the cold room, the frozen droplet was mounted on the specimen holder of a cryo-microtome (SM2010 R, Leica Biosystems, Germany) without embedding medium to avoid spurious Raman signals.
The specimen and chamber were held at \SI{-10}{\degreeCelsius}, and the droplet was trimmed from its apex in steps of \SI{5}{\micro\meter} until a flat horizontal cross-section of radius \(\sim\)\SI{1}{\milli\meter} was exposed.
The freshly exposed section was immediately transferred onto a pre-cooled temperature-controlled stage attached to our confocal Raman microscope (inVia Qontor, Renishaw, UK) to minimize thawing and frost formation.

We acquired the Raman spectra and chemical maps from both the droplet interior and the near-surface layer under cryogenic handling conditions similar to those used in recent frozen-sample studies \cite{Kim2026ArseniteOxidation,Garncarzova2025PhenolFrozen}.
A \SI{532}{\nano\meter} laser, a 2400 lines mm\(^{-1}\) grating, and a long-working-distance \(\times 50\) objective (NA 0.50) provided a lateral probe size of approximately \SI{1}{\micro\meter} and a typical acquisition time of \SI{1}{\second} for static spectra and \SI{0.1}{\second} for chemical mapping.

\begin{figure}[t]
\centering
\includegraphics[width=0.5\linewidth]{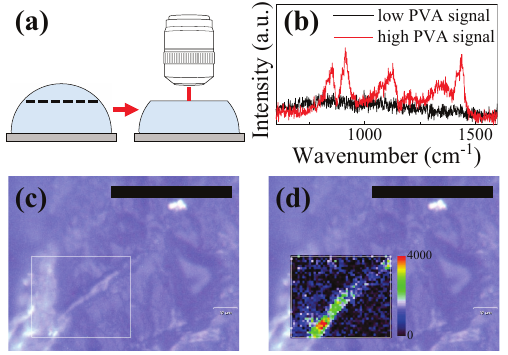}
\caption{Raman characterization of a frozen 3~wt\% PVA solution droplet ($M_\mathrm{w}$~13--23~kDa). (a) Schematic illustration of the preparation and imaging of a horizontal cross-section of a frozen PVA droplet by microtoming. (b) Representative Raman spectra acquired from regions with relatively low and high PVA signal in the horizontal cross-section. (c) Reflection optical image of the cross-sectioned frozen PVA solution droplet (scale bars, \SI{50}{\micro \meter}). (d) Overlaid intensity map of the PVA Raman signal at 1450~cm$^{-1}$ for the region shown in (c). The color bar in the inset indicates Raman intensity in arbitrary units.}
\label{fig_Raman}
\end{figure}

As shown in Fig.~\ref{fig_Raman}(d), we constructed the chemical maps of PVA distribution from the Raman band near \SI{1450}{\per\centi\meter}.
This band, used as a fingerprint of PVA, corresponds to the C--H vibration of PVA~\cite{Hamciuc2020PVARamanPeak}---verified by comparison with spectra acquired from the incompletely frozen PVA solution.
To generate the false-color maps of PVA distribution in Fig.~\ref{fig_Raman}(d), we extracted the intensity in this spectral window at each pixel.
The same procedure was applied to droplets with volumes of \SI{5}{\micro\liter} and \SI{10}{\micro\liter} and to PVA samples with molecular weights of 13--23\,kDa and 85--124\,kDa.

\section{Results and Discussion}

When a sessile droplet of an aqueous polymer solution freezes on a cold substrate, it can develop a rounded apex depending on the polymer concentration~\cite{Kharal2024PolymerDroplets}, instead of the sharp tip singularity of a pure-water droplet~\cite{Marin2014TipSingularity, Im2026XrayMicroscopyDraft}.
Our optical observations reproduce this: increasing the polymer concentration progressively increases the total freezing time and the tip angle of the singularity---eventually reaching \ang{180}, i.e., the rounded apex. 
This phenomenon occurs across all the water-soluble polymers we investigated, i.e., polyvinyl alcohol, polyethylene glycol, polyvinylpyrrolidone, and dextran; see Fig. S1 in the Supplementary Information.
Despite previous studies providing a useful framework for describing how the shape and freezing time depend on the polymer concentration~\cite{Kharal2024PolymerDroplets,Ulrich2026UniversalScalingPreprint}, the following questions remain open: where the polymer rejected from the ice front accumulates during the freezing process, and how its distribution affects the shape and freezing time.

\begin{figure*}[t]
\centering
\includegraphics[width=\linewidth]{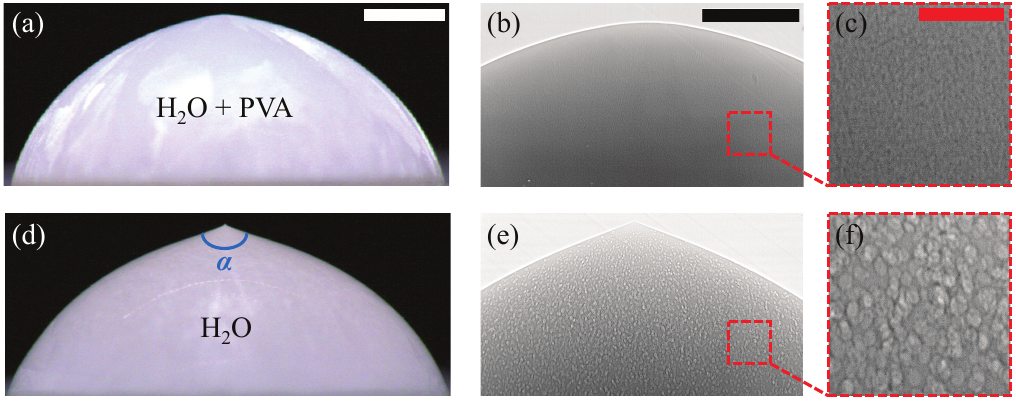}
\caption{Comparison of optical and X-ray observations of frozen water droplets with and without PVA. Optical micrographs of \SI{10}{\micro \liter} frozen sessile droplets acquired in reflection mode under visible-light illumination: (a) aqueous solution of 3~wt\% PVA ($M_\mathrm{w}$~13--23~kDa) and (d) neat water. The white scale bar indicates 1~mm, and the sharp apex in (d) has the tip angle $\alpha$. X-ray radiographs of the upper portions of frozen droplets: (b) aqueous solution of 3~wt\% PVA ($M_\mathrm{w}$~13--23~kDa) and (e) neat water. The black scale bar in (b) indicates \SI{500}{\micro \meter}. (c, f) Enlarged views of the red dashed squares in (b) and (e), respectively, with the \SI{100}{\micro \meter} red scale bar in (c). Air bubbles trapped in ice---round objects with high X-ray transmittance---are shown overlapped in (f), a 2D projected X-ray radiograph.}
\label{fig_OpticsAndRadiograph}
\end{figure*}

\begin{figure}[t]
\centering
\includegraphics[width=0.5\linewidth]{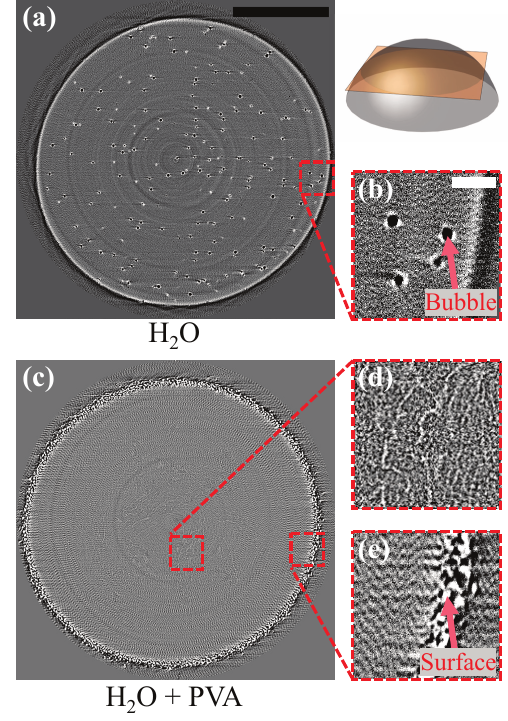}
\caption{Frozen droplets' horizontal cross-sections obtained by X-ray computed tomography: (a) neat water and (c) aqueous solution of 3~wt\% PVA ($M_\mathrm{w}$~13--23~kDa). The black scale bar in (a) is \SI{500}{\micro \meter}. The schematic on the right depicts the cross-sectioning of the droplet. (b) Magnified view of the red box in (a). The dark, round objects show the trapped bubbles; with the inverted lookup table, dark indicates high X-ray transmittance. The white scale bar is \SI{50}{\micro \meter}. The magnified views of the frozen PVA solution droplet in (c) show distinct internal heterogeneity in (d), instead of the bubbles in (b), and rough surface features in (e), instead of the smooth air--ice interface in (b). The overall contrast in (c) is enhanced to better visualize the above-mentioned features.}
\label{fig_HorCrossSection}
\end{figure}

We employ synchrotron X-ray imaging to resolve the internal structure and dynamics of freezing droplets (Fig.~\ref{fig_OpticsAndRadiograph}), which visible light cannot capture due to refraction at the droplets' curved interfaces and scattering from polycrystalline ice.
In the pure-water case shown in Fig.~\ref{fig_OpticsAndRadiograph}(e,f), dissolved gas rejected from the solidifying interface leads to bubble nucleation, growth, and encapsulation in ice, producing discrete pores as the front advances \cite{Meijer2024BubbleGrowth,Thievenaz2025BubbleShape,Im2026XrayMicroscopyDraft}.
Adding polymer---demonstrated here with polyvinyl alcohol (PVA, $M_\mathrm{w}$ 13--23\,kDa) as a representative case---markedly reduces this bubble encapsulation.
As the PVA concentration increases, the number of encapsulated bubbles decreases; above 1~wt\% PVA, we observe no discrete bubbles at our X-ray imaging resolution of \SI{0.65}{\micro\meter} (Fig.~\ref{fig_OpticsAndRadiograph}(b,c), Fig. SF4).
As shown in Fig.~\ref{fig_HorCrossSection}(c,d), tomographic reconstruction further demonstrates that frozen high-concentration PVA droplets do not contain bubbles but rather interconnected low-transmittance domains, with a width of the order of \SI{5}{\micro\meter}, distributed throughout the interior.
We consistently observe both tip blunting and suppression of discrete bubble entrapment across the investigated droplet volume range of \SIrange{1}{10}{\micro\liter} (Fig. S2).

Raman spectral mapping of the horizontal cross-sections of the frozen PVA droplets reveals how this structural heterogeneity correlates with polymer distribution.
Fig.~\ref{fig_Raman} suggests that the interconnected low-transmittance regions identified by tomography correspond to domains with enhanced PVA Raman signal.
The chemical maps are constructed from the Raman peak near \SI{1450}{\per\centi\meter}, which serves as a fingerprint of PVA.
Comparison with spectra acquired from the `incompletely frozen' PVA solution confirms that the C--H vibration of PVA is responsible for this band, indicating that these regions correspond to PVA-enriched incompletely frozen domains rather than simple density variations.
We presume that dissolved gas can remain in these incompletely frozen domains rather than being rejected and trapped as bubbles in ice.
In summary, X-ray tomography and Raman microscopy indicate that PVA rejected from the growing ice matrix accumulates within incompletely frozen domains as the freezing front advances.

\begin{figure}[t]
\centering
\includegraphics[width=0.5\linewidth]{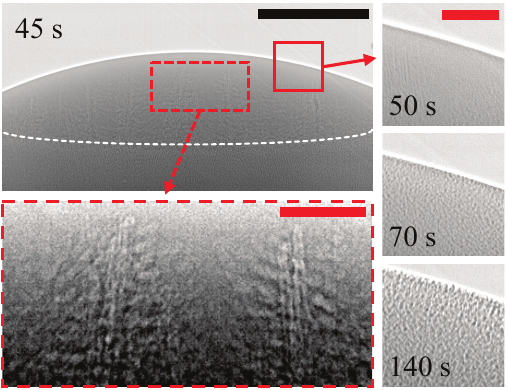}
\caption{Time-resolved X-ray radiographs of the freezing process. The upper part of a \SI{10}{\micro \liter} sessile droplet of 3~wt\% PVA ($M_\mathrm{w}$~13--23~kDa) is shown in the upper left panel, with the black scale bar indicating \SI{500}{\micro \meter}. The white dashed line marks the advancing freezing front. The magnified, contrast-enhanced view of the red-dashed box is shown below, with the red scale bar indicating \SI{100}{\micro \meter}. The right panels enlarge the red solid box and show how the droplet surface develops a bumpy texture over time, with the red scale bar indicating \SI{100}{\micro \meter}.}
\label{fig_BurrAndBump}
\end{figure}

A close look at the freezing front in the time-resolved X-ray radiography hints at how polymer redistribution occurs.
As shown in Fig.~\ref{fig_BurrAndBump}, we observe a burr-like solid--liquid interface that becomes rougher and propagates more slowly as PVA concentration increases, rather than the smooth freezing front observed in pure-water droplets.
Such roughening is the established signature of solute rejection at a moving solidification front~\cite{Mullins1963MorphologicalStability,Mullins1964PlanarInterface,Zhang2021PlanarInstability}. 
When rejected solutes cannot diffuse away rapidly enough to homogenize the concentration field ahead of the interface, the resulting gradient can generate constitutional supercooling and destabilize an initially planar front.
Similar burr-like fronts have been reported in saltwater freezing~\cite{Zhu2025SaltwaterBurrLike}, resulting in the formation of brine pockets and channels, which resemble the polymer-enriched domains. 
Additionally, the same transport-limited regime has been argued to govern droplet-scale morphology across a wide range of polymer solutions \cite{Ulrich2026UniversalScalingPreprint}.
Namely, the frozen droplet of an aqueous polymer solution is better described as a spatially heterogeneous solid embedded with incompletely frozen domains shaped by polymer exclusion and freeze concentration than as a uniform ice matrix. 

Our findings provide a spatially resolved interpretation of the tip blunting observed in Fig.~\ref{fig_OpticsAndRadiograph}. 
As the solid-to-liquid density ratio $\nu$ approaches 1, the angle $\alpha$ of the tip singularity increases, eventually reaching \ang{180}, i.e., the rounded apex~\cite{Marin2014TipSingularity,Kharal2024PolymerDroplets}. 
Our observations imply that $\nu$ of polymer solutions is no longer a uniform bulk property.
It is better interpreted as a local property of a heterogeneous solid embedded with polymer-rich incompletely frozen domains.
The more incompletely frozen domains there are near the apex, the closer the effective $\nu$ gets to 1, leading to tip blunting.
Similarly, a reduction in the effective $\nu$ due to bubble entrapment can result in tip sharpening~\cite{Im2026XrayMicroscopyDraft}.

The same interpretation applies to the increase in total freezing time with PVA concentration.
In the bulk description, the freezing time $\tau$ follows the quasi-steady unidirectional solidification expression $\tau = \frac{\rho_f L H_f^2}{2\lambda_f \Delta T}$, where $\rho_f$ is the frozen-phase density, $L$ is the latent heat, $H_f$ is the final frozen height, $\lambda_f$ is the thermal conductivity, and $\Delta T$ is the temperature difference between the solution and the cold substrate \cite{Wang2022FreezingTime,Zhang2017FreezingTime,Tembely2019FreezingTime,Kharal2024PolymerDroplets}.
Because the solidifying body is not homogeneous but a physicochemical composite of polymer-enriched domains and surrounding ice, a more complete description of the freezing-time trend must account for this spatial inhomogeneity, of which the burr-like front is a directly observable manifestation.

Lastly, we report that the polymer redistribution process also produces a distinct change in outer-surface morphology.
As shown in Fig.~\ref{fig_OpticsAndRadiograph}(a,d), frozen PVA droplets seem to scatter more light than pure-water droplets in optical reflection images under identical illumination conditions.
X-ray radiography in Fig.~\ref{fig_BurrAndBump} captures the development of the outer-surface structure during freezing.
Tomographic cross-sections in Figs.~\ref{fig_HorCrossSection}(e) and S3 also reveal the rough outer layer with lower transmittance than the ice interior, composed of outward protrusions oriented approximately normal to the local interface, and this roughness is already detectable at low PVA concentrations as shown in Fig. S4(b).
Raman mapping in Fig.~\ref{fig_Raman} further shows enhanced PVA signal in this outer layer, indicating that the protrusions are associated with polymer enrichment rather than with surface topography alone.
These observations are consistent with a transport-limited freezing process, in which rejected polymer is not homogenized during front advance but instead accumulates in spatially distinct regions, including both the droplet interior and the near-surface layer.
Whether this outer layer and the interior domains originate from the same formation mechanism and by what pathway the polymer accumulates near the surface warrant future investigation.

\section{Conclusion}

Aqueous PVA droplets freeze through a spatially heterogeneous solidification pathway that blunts the singular tip and reduces discrete bubble entrapment relative to pure water.
Optical imaging, synchrotron X-ray imaging, and Raman spectroscopy collectively show that freezing is accompanied by PVA redistribution, which roughens the advancing front and produces both a polymer-enriched surface layer and interior channel-like domains.
Our results identify freeze-induced polymer segregation as the spatial basis underlying both tip blunting and reduced bubble entrapment, complementing recent macroscopic scaling descriptions of freezing polymer droplets \cite{Ulrich2026UniversalScalingPreprint} by revealing the spatial heterogeneity that underlies the averaged morphological response.

The following questions warrant further investigation: how the polymer-rich domains form, what phase they are in, and where the rejected gas goes.
Additionally, whether the morphology and abundance of those domains depend systematically on polymer identity (e.g., ice-binding versus non-ice-binding polymers), molecular weight, and freezing rate also remains to be tested, and would clarify the generality of freeze-induced segregation in shaping frozen droplets.
These questions are relevant to freezing-based processes such as freeze-casting and cryopreservation, where spatially resolved characterization of solute redistribution can help guide process design.

\printcredits

\section*{Data availability}
Additional data, such as raw images and data, can be obtained from \cite{Zenodo}. Further data will be made available on request.

\section*{Declaration of competing interest}
The authors declare that they have no known competing financial interests or personal relationships that could have appeared to influence the work reported in this paper.

\section*{Acknowledgement}
This work was supported by the Korea Polar Research Institute (KOPRI) grant funded by the Ministry of Oceans and Fisheries (KOPRI PE25900, PE26120). The authors also acknowledge partial financial support from the Korean National Research Foundation through Grant No. RS-2024-00437443. Experiments using PLS-II (6C BMI beamline) were supported in part by the Ministry of Science and ICT (MSIT, RS-2022-00164805, Accelerator Application Support Project) and POSTECH. We thank the UNIST Office of Research Facilities and Training (ResFacT) for support with the use of the wire-cut EDM equipment (SL400G, Sodick, Japan).

\section*{Supplementary material}
Supplementary material is available in the online version of the paper.
 

\begin{thebibliography}{43}
\expandafter\ifx\csname natexlab\endcsname\relax\def\natexlab#1{#1}\fi
\providecommand{\url}[1]{\texttt{#1}}
\providecommand{\href}[2]{#2}
\providecommand{\path}[1]{#1}
\providecommand{\DOIprefix}{doi:}
\providecommand{\ArXivprefix}{arXiv:}
\providecommand{\URLprefix}{URL: }
\providecommand{\Pubmedprefix}{pmid:}
\providecommand{\doi}[1]{\href{http://dx.doi.org/#1}{\path{#1}}}
\providecommand{\Pubmed}[1]{\href{pmid:#1}{\path{#1}}}
\providecommand{\bibinfo}[2]{#2}
\ifx\xfnm\relax \def\xfnm[#1]{\unskip,\space#1}\fi
\bibitem[{Chu et~al.(2024)Chu, Li, Zhao, Feng, Lin, Wu, Yan, and Miljkovic}]{Chu2024SaltwaterNatComm}
\bibinfo{author}{F.~Chu}, \bibinfo{author}{S.~Li}, \bibinfo{author}{C.~Zhao}, \bibinfo{author}{Y.~Feng}, \bibinfo{author}{Y.~Lin}, \bibinfo{author}{X.~Wu}, \bibinfo{author}{X.~Yan}, \bibinfo{author}{N.~Miljkovic},
\newblock \bibinfo{title}{Interfacial ice sprouting during salty water droplet freezing},
\newblock \bibinfo{journal}{Nature Communications} \bibinfo{volume}{15} (\bibinfo{year}{2024}) \bibinfo{pages}{2249}.
\bibitem[{Zhu et~al.(2025)Zhu, Wang, Dai, Wang, and Wang}]{Zhu2025SaltwaterBurrLike}
\bibinfo{author}{J.~Zhu}, \bibinfo{author}{Z.~Wang}, \bibinfo{author}{Z.~Dai}, \bibinfo{author}{Y.~Wang}, \bibinfo{author}{M.~Wang},
\newblock \bibinfo{title}{Burr-like structures growth and diffuse freezing front during saltwater droplet impact freezing},
\newblock \bibinfo{journal}{Journal of Colloid and Interface Science} \bibinfo{volume}{699} (\bibinfo{year}{2025}) \bibinfo{pages}{138227}.
\bibitem[{Wang et~al.(2024)Wang, Chen, Li, Huo, Gu, Hu, and Deng}]{Wang2024SelfLiftingPRL}
\bibinfo{author}{F.~Wang}, \bibinfo{author}{L.~Chen}, \bibinfo{author}{Y.~Li}, \bibinfo{author}{P.~Huo}, \bibinfo{author}{X.~Gu}, \bibinfo{author}{M.~Hu}, \bibinfo{author}{D.~Deng},
\newblock \bibinfo{title}{Self-lifting droplet driven by the solidification-induced solutal marangoni flow},
\newblock \bibinfo{journal}{Physical Review Letters} \bibinfo{volume}{132} (\bibinfo{year}{2024}) \bibinfo{pages}{014002}.
\bibitem[{Shen et~al.(2025)Shen, Fang, Zhang, Zhang, and Tao}]{Shen2025EthanolDropletBulge}
\bibinfo{author}{F.~Shen}, \bibinfo{author}{W.-Z. Fang}, \bibinfo{author}{S.~Zhang}, \bibinfo{author}{D.~Zhang}, \bibinfo{author}{W.-Q. Tao},
\newblock \bibinfo{title}{Freezing patterns of supercooled binary droplets on cold hydrophobic surfaces},
\newblock \bibinfo{journal}{Langmuir} \bibinfo{volume}{41} (\bibinfo{year}{2025}) \bibinfo{pages}{26705--26714}.
\bibitem[{Snoeijer and Brunet(2012)}]{Snoeijer2012PointyIceDrops}
\bibinfo{author}{J.~H. Snoeijer}, \bibinfo{author}{P.~Brunet},
\newblock \bibinfo{title}{Pointy ice-drops: How water freezes into a singular shape},
\newblock \bibinfo{journal}{American Journal of Physics} \bibinfo{volume}{80} (\bibinfo{year}{2012}) \bibinfo{pages}{764--771}.
\bibitem[{Marin et~al.(2014)Marin, Enriquez, Brunet, Colinet, and Snoeijer}]{Marin2014TipSingularity}
\bibinfo{author}{A.~G. Marin}, \bibinfo{author}{O.~R. Enriquez}, \bibinfo{author}{P.~Brunet}, \bibinfo{author}{P.~Colinet}, \bibinfo{author}{J.~H. Snoeijer},
\newblock \bibinfo{title}{Universality of tip singularity formation in freezing water drops},
\newblock \bibinfo{journal}{Physical Review Letters} \bibinfo{volume}{113} (\bibinfo{year}{2014}) \bibinfo{pages}{054301}.
\bibitem[{Chu et~al.(2019)Chu, Zhang, Li, Jin, Zhang, Wu, and Wen}]{Chu2019BubbleFormation}
\bibinfo{author}{F.~Chu}, \bibinfo{author}{X.~Zhang}, \bibinfo{author}{S.~Li}, \bibinfo{author}{H.~Jin}, \bibinfo{author}{J.~Zhang}, \bibinfo{author}{X.~Wu}, \bibinfo{author}{D.~Wen},
\newblock \bibinfo{title}{Bubble formation in freezing droplets},
\newblock \bibinfo{journal}{Physical Review Fluids} \bibinfo{volume}{4} (\bibinfo{year}{2019}) \bibinfo{pages}{071601}.
\bibitem[{Li et~al.(2022)Li, Li, Dang, and Liu}]{Li2022DissolvedGas}
\bibinfo{author}{Y.~Li}, \bibinfo{author}{M.~Li}, \bibinfo{author}{C.~Dang}, \bibinfo{author}{X.~Liu},
\newblock \bibinfo{title}{Effects of dissolved gas on the nucleation and growth of ice crystals in freezing droplets},
\newblock \bibinfo{journal}{International Journal of Heat and Mass Transfer} \bibinfo{volume}{184} (\bibinfo{year}{2022}) \bibinfo{pages}{122334}.
\bibitem[{Meijer et~al.(2024)Meijer, Rocha, Linnenbank, Diddens, and Lohse}]{Meijer2024BubbleGrowth}
\bibinfo{author}{J.~G. Meijer}, \bibinfo{author}{D.~Rocha}, \bibinfo{author}{A.~M. Linnenbank}, \bibinfo{author}{C.~Diddens}, \bibinfo{author}{D.~Lohse},
\newblock \bibinfo{title}{Enhanced bubble growth near an advancing solidification front},
\newblock \bibinfo{journal}{Journal of Fluid Mechanics} \bibinfo{volume}{996} (\bibinfo{year}{2024}) \bibinfo{pages}{A22}.
\bibitem[{Shao et~al.(2024)Shao, Song, Shen, Zhang, and Peka\v{r}}]{Shao2024BubbleDistribution}
\bibinfo{author}{K.~Shao}, \bibinfo{author}{M.~Song}, \bibinfo{author}{J.~Shen}, \bibinfo{author}{X.~Zhang}, \bibinfo{author}{L.~Peka\v{r}},
\newblock \bibinfo{title}{Experimental study on the distribution and growth characteristics of trapped air bubbles in ice slices at different freezing temperatures},
\newblock \bibinfo{journal}{Applied Thermal Engineering} \bibinfo{volume}{244} (\bibinfo{year}{2024}) \bibinfo{pages}{122600}.
\bibitem[{Shao et~al.(2023)Shao, Song, Zhang, and Zhang}]{Shao2023BubbleGrowthDistribution}
\bibinfo{author}{K.~Shao}, \bibinfo{author}{M.~Song}, \bibinfo{author}{X.~Zhang}, \bibinfo{author}{L.~Zhang},
\newblock \bibinfo{title}{Growth and distribution characteristics of trapped air bubbles in ice slices},
\newblock \bibinfo{journal}{Physics of Fluids} \bibinfo{volume}{35} (\bibinfo{year}{2023}) \bibinfo{pages}{113319}.
\bibitem[{Snels et~al.(2023)Snels, Sarkari, Soete, Maes, Antonini, Wevers, Maitra, and Seveno}]{Snels2023IceDropletXCT}
\bibinfo{author}{L.~Snels}, \bibinfo{author}{N.~M. Sarkari}, \bibinfo{author}{J.~Soete}, \bibinfo{author}{A.~Maes}, \bibinfo{author}{C.~Antonini}, \bibinfo{author}{M.~Wevers}, \bibinfo{author}{T.~Maitra}, \bibinfo{author}{D.~Seveno},
\newblock \bibinfo{title}{Internal and interfacial microstructure characterization of ice droplets on surfaces by {X}-ray computed tomography},
\newblock \bibinfo{journal}{Journal of Colloid and Interface Science} \bibinfo{volume}{637} (\bibinfo{year}{2023}) \bibinfo{pages}{500--512}.
\bibitem[{Thi\'{e}venaz et~al.(2025)Thi\'{e}venaz, Meijer, Lohse, and Sauret}]{Thievenaz2025BubbleShape}
\bibinfo{author}{V.~Thi\'{e}venaz}, \bibinfo{author}{J.~G. Meijer}, \bibinfo{author}{D.~Lohse}, \bibinfo{author}{A.~Sauret},
\newblock \bibinfo{title}{On the shape of air bubbles trapped in ice},
\newblock \bibinfo{journal}{Proceedings of the National Academy of Sciences} \bibinfo{volume}{122} (\bibinfo{year}{2025}) \bibinfo{pages}{e2415027122}.
\bibitem[{Im et~al.(2025)Im, An, Kim, Lim, and Jeong}]{Im2026XrayMicroscopyDraft}
\bibinfo{author}{J.~K. Im}, \bibinfo{author}{H.~An}, \bibinfo{author}{S.-G. Kim}, \bibinfo{author}{J.-H. Lim}, \bibinfo{author}{J.~Jeong},
\newblock \bibinfo{title}{{X}-ray microscopy study of freezing sessile droplets},
\newblock \bibinfo{journal}{arXiv preprint}  (\bibinfo{year}{2025}).
\bibitem[{Budke and Koop(2006)}]{Budke2006PVAIRI}
\bibinfo{author}{C.~Budke}, \bibinfo{author}{T.~Koop},
\newblock \bibinfo{title}{Ice recrystallization inhibition and molecular recognition of ice faces by poly (vinyl alcohol)},
\newblock \bibinfo{journal}{ChemPhysChem} \bibinfo{volume}{7} (\bibinfo{year}{2006}) \bibinfo{pages}{2601--2606}.
\bibitem[{Naullage et~al.(2018)Naullage, Qiu, and Molinero}]{Naullage2018PinnedIceSurfaces}
\bibinfo{author}{P.~M. Naullage}, \bibinfo{author}{Y.~Qiu}, \bibinfo{author}{V.~Molinero},
\newblock \bibinfo{title}{What controls the limit of supercooling and superheating of pinned ice surfaces?},
\newblock \bibinfo{journal}{The Journal of Physical Chemistry Letters} \bibinfo{volume}{9} (\bibinfo{year}{2018}) \bibinfo{pages}{1712--1720}.
\bibitem[{Naullage and Molinero(2020)}]{Naullage2020PVAIRI}
\bibinfo{author}{P.~M. Naullage}, \bibinfo{author}{V.~Molinero},
\newblock \bibinfo{title}{Slow propagation of ice binding limits the ice-recrystallization inhibition efficiency of {PVA} and other flexible polymers},
\newblock \bibinfo{journal}{Journal of the American Chemical Society} \bibinfo{volume}{142} (\bibinfo{year}{2020}) \bibinfo{pages}{4356--4366}.
\bibitem[{Bachtiger et~al.(2021)Bachtiger, Congdon, Stubbs, Gibson, and Sosso}]{Bachtiger2021PVAIRI}
\bibinfo{author}{F.~Bachtiger}, \bibinfo{author}{T.~R. Congdon}, \bibinfo{author}{C.~Stubbs}, \bibinfo{author}{M.~I. Gibson}, \bibinfo{author}{G.~C. Sosso},
\newblock \bibinfo{title}{The atomistic details of the ice recrystallisation inhibition activity of {PVA}},
\newblock \bibinfo{journal}{Nature Communications} \bibinfo{volume}{12} (\bibinfo{year}{2021}) \bibinfo{pages}{1323}.
\bibitem[{Weng et~al.(2018)Weng, Stott, and Toner}]{Weng2018PVAIceInterface}
\bibinfo{author}{L.~Weng}, \bibinfo{author}{S.~L. Stott}, \bibinfo{author}{M.~Toner},
\newblock \bibinfo{title}{Molecular dynamics at the interface between ice and poly(vinyl alcohol) and ice recrystallization inhibition},
\newblock \bibinfo{journal}{Langmuir} \bibinfo{volume}{34} (\bibinfo{year}{2018}) \bibinfo{pages}{5116--5123}.
\bibitem[{Zhang et~al.(2021{\natexlab{a}})Zhang, Wang, Wang, Li, and Wang}]{Zhang2021SingleIceCrystal}
\bibinfo{author}{T.~Zhang}, \bibinfo{author}{L.~Wang}, \bibinfo{author}{Z.~Wang}, \bibinfo{author}{J.~Li}, \bibinfo{author}{J.~Wang},
\newblock \bibinfo{title}{Single ice crystal growth with controlled orientation during directional freezing},
\newblock \bibinfo{journal}{The Journal of Physical Chemistry B} \bibinfo{volume}{125} (\bibinfo{year}{2021}{\natexlab{a}}) \bibinfo{pages}{970--979}.
\bibitem[{Zhang et~al.(2021{\natexlab{b}})Zhang, Wang, Wang, Li, and Wang}]{Zhang2021Lamellar}
\bibinfo{author}{T.~Zhang}, \bibinfo{author}{Z.~Wang}, \bibinfo{author}{L.~Wang}, \bibinfo{author}{J.~Li}, \bibinfo{author}{J.~Wang},
\newblock \bibinfo{title}{Tilting behavior of lamellar ice tip during unidirectional freezing of aqueous solutions},
\newblock \bibinfo{journal}{Langmuir} \bibinfo{volume}{37} (\bibinfo{year}{2021}{\natexlab{b}}) \bibinfo{pages}{10579--10587}.
\bibitem[{Piao et~al.(2021)Piao, Park, Patel, Lee, and Jeong}]{Piao2021PVAIRIneedleShape}
\bibinfo{author}{Z.~Piao}, \bibinfo{author}{J.~K. Park}, \bibinfo{author}{M.~Patel}, \bibinfo{author}{H.~J. Lee}, \bibinfo{author}{B.~Jeong},
\newblock \bibinfo{title}{Poly ({L}-ala-co-{L}-lys) exhibits excellent ice recrystallization inhibition activity},
\newblock \bibinfo{journal}{ACS Macro Letters} \bibinfo{volume}{10} (\bibinfo{year}{2021}) \bibinfo{pages}{1436--1442}.
\bibitem[{Kharal and Louf(2024)}]{Kharal2024PolymerDroplets}
\bibinfo{author}{S.~P. Kharal}, \bibinfo{author}{J.-F. Louf},
\newblock \bibinfo{title}{Unidirectional freezing of polymer solution droplets},
\newblock \bibinfo{journal}{Langmuir} \bibinfo{volume}{40} (\bibinfo{year}{2024}) \bibinfo{pages}{118--124}.
\bibitem[{Ulrich and Louf(2026)}]{Ulrich2026PolymerFreezingReview}
\bibinfo{author}{N.~G. Ulrich}, \bibinfo{author}{J.-F. Louf},
\newblock \bibinfo{title}{Interfacial mechanisms in the freezing of polymer solutions},
\newblock \bibinfo{journal}{Nanoscale} \bibinfo{volume}{18} (\bibinfo{year}{2026}) \bibinfo{pages}{3453--3471}.
\bibitem[{Wegst et~al.(2010)Wegst, Schecter, Donius, and Hunger}]{Wegst2010BiomaterialsFreezeCasting}
\bibinfo{author}{U.~G.~K. Wegst}, \bibinfo{author}{M.~Schecter}, \bibinfo{author}{A.~E. Donius}, \bibinfo{author}{P.~M. Hunger},
\newblock \bibinfo{title}{Biomaterials by freeze casting},
\newblock \bibinfo{journal}{Philosophical Transactions of the Royal Society A: Mathematical, Physical and Engineering Sciences} \bibinfo{volume}{368} (\bibinfo{year}{2010}) \bibinfo{pages}{2099--2121}.
\bibitem[{Yin et~al.(2019)Yin, Divakar, and Wegst}]{Yin2019FreezeCastChitosan}
\bibinfo{author}{K.~Yin}, \bibinfo{author}{P.~Divakar}, \bibinfo{author}{U.~G.~K. Wegst},
\newblock \bibinfo{title}{Plant-derived nanocellulose as structural and mechanical reinforcement of freeze-cast chitosan scaffolds for biomedical applications},
\newblock \bibinfo{journal}{Biomacromolecules} \bibinfo{volume}{20} (\bibinfo{year}{2019}) \bibinfo{pages}{3733--3745}.
\bibitem[{Yin et~al.(2023)Yin, Ji, Littles, Trivedi, Karma, and Wegst}]{Yin2023PolymerFreezeCasting}
\bibinfo{author}{K.~Yin}, \bibinfo{author}{K.~Ji}, \bibinfo{author}{L.~S. Littles}, \bibinfo{author}{R.~Trivedi}, \bibinfo{author}{A.~Karma}, \bibinfo{author}{U.~G.~K. Wegst},
\newblock \bibinfo{title}{Hierarchical structure formation by crystal growth-front instabilities during ice templating},
\newblock \bibinfo{journal}{Proceedings of the National Academy of Sciences} \bibinfo{volume}{120} (\bibinfo{year}{2023}) \bibinfo{pages}{e2210242120}.
\bibitem[{Ulrich et~al.(2026)Ulrich, Aravindhan, Berger, Beckingham, and Louf}]{Ulrich2026UniversalScalingPreprint}
\bibinfo{author}{N.~G. Ulrich}, \bibinfo{author}{P.~P. Aravindhan}, \bibinfo{author}{O.~Berger}, \bibinfo{author}{B.~S. Beckingham}, \bibinfo{author}{J.-F. Louf},
\newblock \bibinfo{title}{Universal scaling of freezing morphodynamics in polymer solution droplets},
\newblock \bibinfo{journal}{arXiv preprint}  (\bibinfo{year}{2026}).
\bibitem[{Kamm et~al.(2023)Kamm, Yin, Neu, Schlep{\"u}tz, Garc{\'i}a-Moreno, and Wegst}]{Kamm2023XraySucrose}
\bibinfo{author}{P.~H. Kamm}, \bibinfo{author}{K.~Yin}, \bibinfo{author}{T.~R. Neu}, \bibinfo{author}{C.~M. Schlep{\"u}tz}, \bibinfo{author}{F.~Garc{\'i}a-Moreno}, \bibinfo{author}{U.~G.~K. Wegst},
\newblock \bibinfo{title}{{X}-ray tomoscopy reveals the dynamics of ice templating},
\newblock \bibinfo{journal}{Advanced Functional Materials} \bibinfo{volume}{33} (\bibinfo{year}{2023}) \bibinfo{pages}{2304738}.
\bibitem[{Alpert et~al.(2022)Alpert, Boucly, Yang, Yang, Kilchhofer, Luo, Padeste, Finizio, Ammann, and Watts}]{Alpert2022XrayIceNucleation}
\bibinfo{author}{P.~A. Alpert}, \bibinfo{author}{A.~Boucly}, \bibinfo{author}{S.~Yang}, \bibinfo{author}{H.~Yang}, \bibinfo{author}{K.~Kilchhofer}, \bibinfo{author}{Z.~Luo}, \bibinfo{author}{C.~Padeste}, \bibinfo{author}{S.~Finizio}, \bibinfo{author}{M.~Ammann}, \bibinfo{author}{B.~Watts},
\newblock \bibinfo{title}{Ice nucleation imaged with {X}-ray spectro-microscopy},
\newblock \bibinfo{journal}{Environmental Science: Atmospheres} \bibinfo{volume}{2} (\bibinfo{year}{2022}) \bibinfo{pages}{335--351}.
\bibitem[{Wu et~al.(2025)Wu, Gao, Jia, Jiang, Huppert, and Lei}]{Wu2025XraySaltwater}
\bibinfo{author}{Z.~Wu}, \bibinfo{author}{X.~Gao}, \bibinfo{author}{J.~Jia}, \bibinfo{author}{Y.~Jiang}, \bibinfo{author}{H.~E. Huppert}, \bibinfo{author}{L.~Lei},
\newblock \bibinfo{title}{Microscopic salt exclusion dynamics of directional freezing in brine},
\newblock \bibinfo{journal}{Journal of Fluid Mechanics} \bibinfo{volume}{1021} (\bibinfo{year}{2025}) \bibinfo{pages}{A15}.
\bibitem[{Jin et~al.(2017)Jin, Yurkow, Adler, and Lee}]{Jin2017XrayINP}
\bibinfo{author}{J.~Jin}, \bibinfo{author}{E.~J. Yurkow}, \bibinfo{author}{D.~Adler}, \bibinfo{author}{T.-C. Lee},
\newblock \bibinfo{title}{A novel approach to improve the efficiency of block freeze concentration using ice nucleation proteins with altered ice morphology},
\newblock \bibinfo{journal}{Journal of Agricultural and Food Chemistry} \bibinfo{volume}{65} (\bibinfo{year}{2017}) \bibinfo{pages}{2373--2382}.
\bibitem[{Seibert et~al.(1998)Seibert, Boone, and Lindfors}]{Seibert1998FlatField}
\bibinfo{author}{J.~A. Seibert}, \bibinfo{author}{J.~M. Boone}, \bibinfo{author}{K.~K. Lindfors},
\newblock \bibinfo{title}{Flat-field correction technique for digital detectors},
\newblock in: \bibinfo{booktitle}{Medical Imaging 1998: Physics of Medical Imaging}, volume \bibinfo{volume}{3336} of \textit{\bibinfo{series}{Proceedings of SPIE}}, \bibinfo{year}{1998}, pp. \bibinfo{pages}{348--354}. \DOIprefix\doi{10.1117/12.317034}.
\bibitem[{An et~al.(2026)An, Kim, Im, Kim, Kim, Lim, Kim, and Jeong}]{Zenodo}
\bibinfo{author}{H.~An}, \bibinfo{author}{B.~Kim}, \bibinfo{author}{J.~K. Im}, \bibinfo{author}{M.~W. Kim}, \bibinfo{author}{S.-G. Kim}, \bibinfo{author}{J.-H. Lim}, \bibinfo{author}{K.~Kim}, \bibinfo{author}{J.~Jeong},
\newblock \bibinfo{title}{Polymer-regulated freezing of water droplets revealed by synchrotron x-ray imaging and raman spectroscopy [dataset]},
\newblock \bibinfo{journal}{ZENODO}  (\bibinfo{year}{2026}).
\bibitem[{Kim et~al.(2026)Kim, Shin, Tran, Ahn, Kim, and Kim}]{Kim2026ArseniteOxidation}
\bibinfo{author}{K.~Kim}, \bibinfo{author}{G.~Shin}, \bibinfo{author}{K.~D. Tran}, \bibinfo{author}{Y.-Y. Ahn}, \bibinfo{author}{B.~Kim}, \bibinfo{author}{J.~Kim},
\newblock \bibinfo{title}{Freezing-induced oxidation of arsenite by nitrite through a chain reaction mechanism},
\newblock \bibinfo{journal}{Separation and Purification Technology} \bibinfo{volume}{382} (\bibinfo{year}{2026}) \bibinfo{pages}{135736}.
\bibitem[{Garncarzov\'{a} et~al.(2025)Garncarzov\'{a}, Vesel\'{y}, Kim, Kim, and Heger}]{Garncarzova2025PhenolFrozen}
\bibinfo{author}{M.~Garncarzov\'{a}}, \bibinfo{author}{L.~Vesel\'{y}}, \bibinfo{author}{B.~Kim}, \bibinfo{author}{K.~Kim}, \bibinfo{author}{D.~Heger},
\newblock \bibinfo{title}{Spectroscopic characterization of phenol in frozen aqueous solution and on the ice surface},
\newblock \bibinfo{journal}{Spectrochimica Acta Part A: Molecular and Biomolecular Spectroscopy} \bibinfo{volume}{335} (\bibinfo{year}{2025}) \bibinfo{pages}{125948}.
\bibitem[{Hamciuc et~al.(2020)Hamciuc, Vlad-Bubulac, Serbezeanu, Hamciuc, Aflori, Lisa, Anghel, \c{S}ofran, and Trofin}]{Hamciuc2020PVARamanPeak}
\bibinfo{author}{C.~Hamciuc}, \bibinfo{author}{T.~Vlad-Bubulac}, \bibinfo{author}{D.~Serbezeanu}, \bibinfo{author}{E.~Hamciuc}, \bibinfo{author}{M.~Aflori}, \bibinfo{author}{G.~Lisa}, \bibinfo{author}{I.~Anghel}, \bibinfo{author}{I.-E. \c{S}ofran}, \bibinfo{author}{A.~Trofin},
\newblock \bibinfo{title}{Tailoring thermal and flame retardant properties via synergistic effect in polyvinyl alcohol nanocomposites based on polyphosphonate and/or sio2 nanoparticles},
\newblock \bibinfo{journal}{Composites Part C: Open Access} \bibinfo{volume}{3} (\bibinfo{year}{2020}) \bibinfo{pages}{100063}.
\bibitem[{Mullins and Sekerka(1963)}]{Mullins1963MorphologicalStability}
\bibinfo{author}{W.~W. Mullins}, \bibinfo{author}{R.~F. Sekerka},
\newblock \bibinfo{title}{Morphological stability of a particle growing by diffusion or heat flow},
\newblock \bibinfo{journal}{Journal of Applied Physics} \bibinfo{volume}{34} (\bibinfo{year}{1963}) \bibinfo{pages}{323--329}.
\bibitem[{Mullins and Sekerka(1964)}]{Mullins1964PlanarInterface}
\bibinfo{author}{W.~W. Mullins}, \bibinfo{author}{R.~F. Sekerka},
\newblock \bibinfo{title}{Stability of a planar interface during solidification of a dilute binary alloy},
\newblock \bibinfo{journal}{Journal of Applied Physics} \bibinfo{volume}{35} (\bibinfo{year}{1964}) \bibinfo{pages}{444--451}.
\bibitem[{Zhang et~al.(2021)Zhang, Wang, Wang, Li, and Wang}]{Zhang2021PlanarInstability}
\bibinfo{author}{T.~Zhang}, \bibinfo{author}{Z.~Wang}, \bibinfo{author}{L.~Wang}, \bibinfo{author}{J.~Li}, \bibinfo{author}{J.~Wang},
\newblock \bibinfo{title}{The planar instability during unidirectional freezing of a macromolecular polymer solution: Diffusion-controlled or not?},
\newblock \bibinfo{journal}{Physica B: Condensed Matter} \bibinfo{volume}{610} (\bibinfo{year}{2021}) \bibinfo{pages}{412923}.
\bibitem[{Wang et~al.(2022)Wang, Xu, Zhang, Zheng, Hao, He, and Zhang}]{Wang2022FreezingTime}
\bibinfo{author}{C.~Wang}, \bibinfo{author}{Z.~Xu}, \bibinfo{author}{H.~Zhang}, \bibinfo{author}{J.~Zheng}, \bibinfo{author}{P.~Hao}, \bibinfo{author}{F.~He}, \bibinfo{author}{X.~Zhang},
\newblock \bibinfo{title}{A new freezing model of sessile droplets considering ice fraction and ice distribution after recalescence},
\newblock \bibinfo{journal}{Physics of Fluids} \bibinfo{volume}{34} (\bibinfo{year}{2022}) \bibinfo{pages}{092115}.
\bibitem[{Zhang et~al.(2017)Zhang, Wu, Min, and Liu}]{Zhang2017FreezingTime}
\bibinfo{author}{X.~Zhang}, \bibinfo{author}{X.~Wu}, \bibinfo{author}{J.~Min}, \bibinfo{author}{X.~Liu},
\newblock \bibinfo{title}{Modelling of sessile water droplet shape evolution during freezing with consideration of supercooling effect},
\newblock \bibinfo{journal}{Applied Thermal Engineering} \bibinfo{volume}{125} (\bibinfo{year}{2017}) \bibinfo{pages}{644--651}.
\bibitem[{Tembely and Dolatabadi(2019)}]{Tembely2019FreezingTime}
\bibinfo{author}{M.~Tembely}, \bibinfo{author}{A.~Dolatabadi},
\newblock \bibinfo{title}{A comprehensive model for predicting droplet freezing features on a cold substrate},
\newblock \bibinfo{journal}{Journal of Fluid Mechanics} \bibinfo{volume}{859} (\bibinfo{year}{2019}) \bibinfo{pages}{566--585}.

\end{thebibliography}

\clearpage

\clearpage

\clearpage

\clearpage

\end{document}


\let\WriteBookmarks\relax
\def\floatpagepagefraction{1}
\def\textpagefraction{.001}
 
\shorttitle{Supplementary Information}
\shortauthors{H. An et~al.}
 
\begin{center}
    {\Large \textbf{Supplementary Information for\\[1ex]
    ``Polymer-Regulated Freezing of Water Droplets Revealed by Synchrotron X-ray Imaging and Raman Spectroscopy''}}\\[2em]
    
    {\large Hyeonjun An$^1$, Bomi Kim$^2$, Jae Kwan Im$^1$, Min Woo Kim$^3$, Seob-Gu Kim$^3$, Jae-Hong Lim$^3$, Kitae Kim$^{2,\ast}$, Joonwoo Jeong$^{1,4,\ast}$}\\[1em]
    
    {\small 
    $^1$ Department of Physics, UNIST, Ulsan, Republic of Korea\\
    $^2$ Korea Polar Research Institute (KOPRI), Incheon, Republic of Korea\\
    $^3$ Pohang Accelerator Laboratory, POSTECH, Pohang, Republic of Korea\\
    $^4$ UNIST Research Center for Soft and Living Matter, UNIST, Ulsan, Republic of Korea\\[1ex]
    $^\ast$ Corresponding authors: ktkim@kopri.re.kr, jjeong@unist.ac.kr
    }
\end{center}
\vspace{2em}

\section*{This PDF file includes:}
\begin{itemize}
\item Figures S1, S2, S3, and S4
\item Captions for Movies S1, S2, S3 and S4
\end{itemize}
 
\section*{Other supplementary materials for this manuscript:}
\begin{itemize}
\item Movies S1, S2, S3 and S4
\end{itemize}
 
\setcounter{figure}{0}
\setcounter{table}{0}
\setcounter{equation}{0}
 
 
\begin{figure}[t]
\centering
\includegraphics[width=\linewidth]{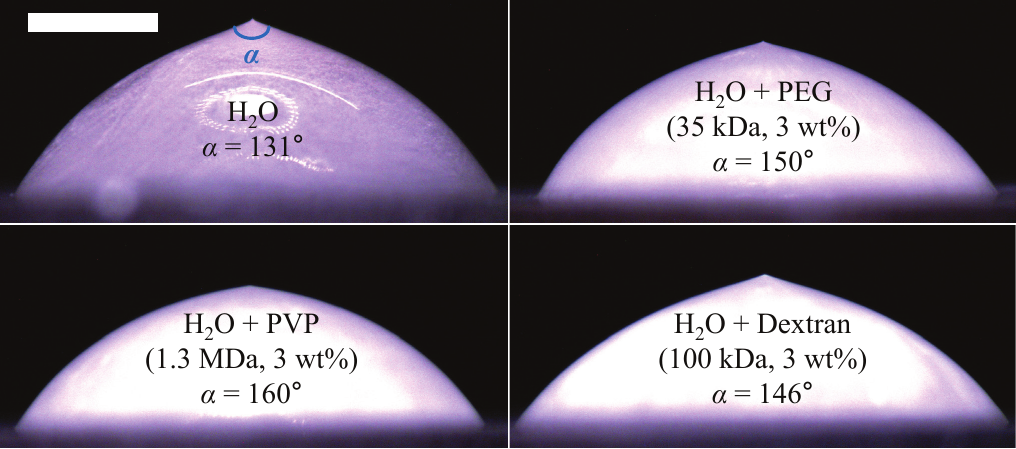}
\caption{Optical micrographs of frozen sessile droplets acquired under reflection mode with visible-light illumination. All droplets have a \SI{10}{\micro\liter} volume and a 3~wt\% polymer concentration. Polyethylene glycol (PEG), polyvinylpyrrolidone (PVP), and dextran all exhibit the same qualitative trends as PVA: upon polymer addition, the tip angle, the total freezing time, and the surface reflectance increase. The white scale bar in the top left panel indicates \SI{1}{\milli\meter}.}
\label{sf1_OtherPolymers}
\end{figure}
 
\begin{figure}[t]
\centering
\includegraphics[width=0.5\linewidth]{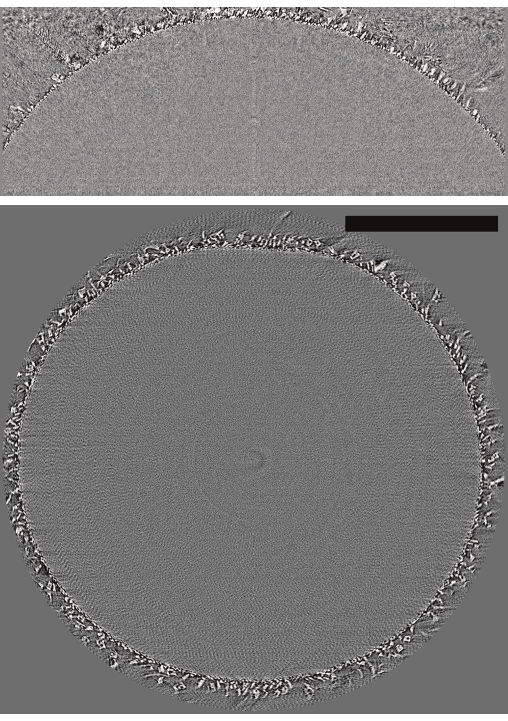}
\caption{Vertical and horizontal cross-sections of the upper part of a frozen \SI{1}{\micro\liter} 3~wt\% PVA ($M_\mathrm{w}$ 13--23\,kDa) solution droplet obtained by X-ray computed tomography. Upon polymer addition, the number of discrete bubbles decreases and the tip angle increases. This trend persists in the \SI{1}{\micro\liter} droplet, which is smaller than the \SI{10}{\micro\liter} droplet used for the representative data in Figure~1 of the main manuscript. The scale bar indicates 500~$\mu$m}
\label{sf2_1uLPVA}
\end{figure}
 
\begin{figure}[t]
\centering
\includegraphics[width=\linewidth]{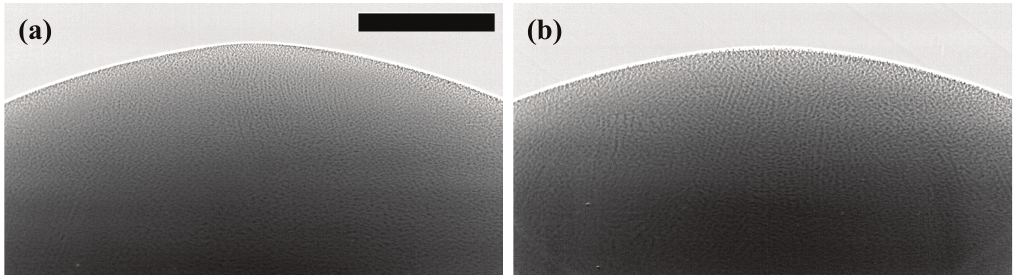}
\caption{X-ray radiographs of the upper portions of frozen \SI{10}{\micro\liter} 3~wt\% PVA ($M_\mathrm{w}$ 85--124\,kDa) solution droplets: (a) typical experimental result; (b) control experiment for X-ray-induced damage, performed with minimized beam exposure during the freezing process. The scale bar indicates 500~$\mu$m}
\label{sf3_XrayDamageControl}
\end{figure}
 
\begin{figure}[t]
\centering 
\includegraphics[width=0.5\linewidth]{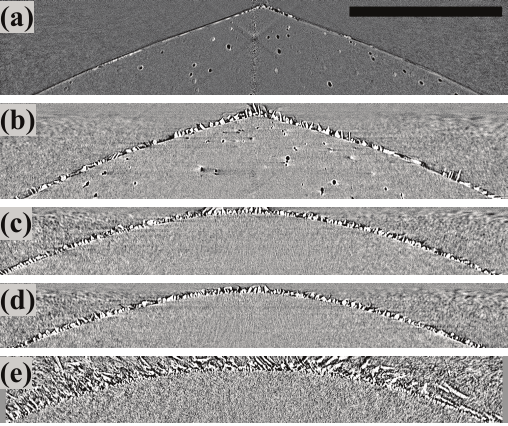}
\caption{Vertical cross-sections of the upper portions of frozen droplets obtained by X-ray computed tomography. (a) Neat water; the black scale bar indicates \SI{500}{\micro\meter}. (b) 0.3~wt\% solution of PVA ($M_\mathrm{w}$ 13--23\,kDa). (c) 1.0~wt\% solution of PVA ($M_\mathrm{w}$ 13--23\,kDa). (d) 3.0~wt\% solution of PVA ($M_\mathrm{w}$ 13--23\,kDa). (e) 3.0~wt\% solution of PVA ($M_\mathrm{w}$ 85--124\,kDa). As the polymer concentration increases, the number of discrete bubbles decreases and the tip angle increases. The same trend holds for the PVA with larger average molecular weight (85--124\,kDa).}
\label{sf4_VerCrossSection}
\end{figure}
 
\clearpage
 
\section*{Captions for Supplementary Movies}
 
\paragraph{Movie S1. X-ray imaging of the freezing dynamics of a PVA solution droplet.}
A \SI{10}{\micro\liter} droplet of 3.0~wt\% PVA ($M_\mathrm{w}$ 13--23\,kDa) was deposited on an aluminum substrate at room temperature, after which the substrate temperature was lowered to \SI{-15}{\degreeCelsius}. Note that no visible encapsulated bubble is observed upon complete solidification. The scale bar indicates \SI{500}{\micro \meter}.
 
\paragraph{Movie S2. Optical imaging of the freezing dynamics of a PVA solution droplet}
A \SI{10}{\micro\liter} droplet of 3.0~wt\% PVA ($M_\mathrm{w}$ 13--23\,kDa) was deposited on an aluminum substrate at room temperature, after which the substrate temperature was lowered to \SI{-15}{\degreeCelsius}. Note that the surface brightness increases after freezing. The scale bar indicates 1~mm.

\paragraph{Movie S3. X-ray imaging of the freezing dynamics of a pure-water droplet.}
A \SI{10}{\micro\liter} pure-water droplet was deposited on an aluminum substrate at room temperature, after which the substrate temperature was lowered to \SI{-15}{\degreeCelsius}. Note that encapsulated bubbles are observed upon complete solidification. The scale bar indicates 500~$\mu$m
 
\paragraph{Movie S4. Optical imaging of the freezing dynamics of a pure-water droplet}
A \SI{10}{\micro\liter} pure-water droplet was deposited on an aluminum substrate at room temperature, after which the substrate temperature was lowered to \SI{-15}{\degreeCelsius}. Note that the surface brightness is lower than that of the PVA solution. The scale bar indicates 1~mm.
 